\documentclass[aip,rsi,reprint]{revtex4-2}
\usepackage{amsmath}
\usepackage{graphicx}
\usepackage{float}
\usepackage{hyperref}
\usepackage{braket}
\usepackage{color}
\begin{document}

\title{Quantum Sensor of Mechanical Vibrations}
\author{Mu Young Kim,  H. C. J. Gan, Nigel Benjamin Lee Junsheng}
\affiliation{Centre for Quantum Technologies, National University of Singapore, 3 Science Dr 2, 117543, Singapore}
\author{Dzmitry Matsukevich}
\affiliation{Centre for Quantum Technologies, National University of Singapore, 3 Science Dr 2, 117543, Singapore}
\affiliation{Department of Physics, National University of Singapore, 2 Science Dr 3, 117551, Singapore}
\date{\today}

\begin{abstract}
We report the use of a single trapped-ion motional mode as a probe of mechanical vibrations of the Paul trap structure. By exciting vibrations of the trap using an ultrasonic transducer mounted outside the vacuum chamber and probing the ion motion through motional sideband interactions, we observe a clear resonance response of the axial mode. We further demonstrate detection of trap vibrations with amplitudes as low as 0.6 picometers. These measurements show direct coupling between macroscopic mechanical motion and trapped-ion dynamics, providing a route toward vibration sensing using quantum control of motional states.
\end{abstract}

\maketitle

\section{Introduction}
Motional modes of trapped ions are a key resource for a wide range of quantum technologies. In quantum information processing, collective modes serve as a quantum bus that mediates interactions between ions~\cite{CiracZoller1995,SorensenMolmer1999a,NIST_Squeezing_2024,bible} in the discrete-variable model of quantum computation. They also provide a harmonic oscillator degree of freedom for continuous-variable quantum computation~\cite{BraunsteinLloydCV,Xiaoguang_Wang_2001} and quantum state engineering~\cite{Fluhmann_gkp_2019,Matsos_GKP_gates_2025}.  In quantum simulation, these modes can be used to determine molecular energy spectra~\cite{Whitlow_Conical_Intersection_2023} and to implement quantum heat machines~\cite{Schmidt_Kaler_Engine_Science,Maslennikov2019}. In quantum metrology, environmental perturbations can be mapped onto changes in motional phase and amplitude, enabling precision sensing of electric and magnetic fields, field gradients, displacements, and forces~\cite{Leibfried2004,Gilmore2017,Degen2017,singer_prl_2023}. Realizing the full potential of these technologies requires understanding and careful characterization of decoherence mechanisms affecting the motional states.

To a good approximation, the motion of an ion in a Paul trap is described by a quantum harmonic oscillator weakly coupled to its environment. Previous work has mainly focused on two types of coupling: heating induced by electric-field noise near trap-electrode surfaces~\cite{Turchette2000,Brownnutt2015,Deslauriers2006}, and motional dephasing caused by fluctuations of the trapping potential~\cite{trap_stabilization_monroe_2016,Oregon_Squaretron_2024}. Coupling of the ion motion to a nanomechanical oscillator integrated in the trap has also been observed \cite{willitsch_prl_2024,willitsch_rsi_2025}.

Another potential mechanism that can affect the motion of trapped ions is mechanical vibrations of the trap itself~\cite{northup_prl_2024,zoller2004}. Although the laboratory environment can often be acoustically  noisy~\cite{schmidt_rsi_2019,pagano_qst_2019}, ultrasound is strongly attenuated in air at frequencies higher than a few hundred kHz~\cite{Ultrasound_Absorption}, which are typical of ion motional modes. In addition, the mechanical resonances of the trap support structure are generally on the order of kilohertz, further suppressing high-frequency oscillations before they reach the trap. Mechanical vibrations of the trap structure near an ion's motional frequency are therefore usually assumed to be negligible. 

Here, we test this assumption experimentally. We show that vibrations of the vacuum chamber, excited by an ultrasonic transducer attached to its exterior, can resonantly drive the ion motion. By cooling the ion to near its motional ground state and probing motional sidebands after driving mechanical vibrations of the vacuum chamber, we characterize this effect and explore the use of trapped ions as quantum sensors of mechanical motion.

We first demonstrate this coupling for the axial mode of a single trapped ion and analyze the resulting dynamics. We then estimate the corresponding trap-displacement amplitude and conclude by discussing the possible use of mechanical driving for motional mode frequency characterization and stabilization.

\section{\label{sec:detection}Setup}
Experiments are performed using a single $^{171}\text{Yb}^+$ ion confined in a four-rod linear Paul trap. Doppler cooling, state initialization, and state detection use the 369 nm transition between $\ket{^2S_{1/2}}$ and $\ket{^2P_{1/2}}$ manifolds~\cite{manipulation_detection}. We denote the hyperfine clock states as $|g\rangle \equiv |F = 0, m_F = 0\rangle$ and $|e\rangle \equiv |F = 1, m_F = 0\rangle$, and coherently control the internal and motional degrees of freedom using stimulated Raman transitions driven by a mode-locked, pulsed $355\,\mathrm{nm}$ laser~\cite{ultrafast}. A counter-propagating beam geometry enables coupling to all three motional modes. The mode frequencies, measured by sideband spectroscopy, are $(\omega_x,\omega_y,\omega_z) / 2\pi = (1.364,\,1.117,\,0.355)$ MHz. The corresponding Lamb-Dicke parameters are $(\eta_x, \eta_y, \eta_z) \approx (0.08, 0.09, 0.26)$ with heating rates $(\Gamma_x, \Gamma_y, \Gamma_z) = (3.7, 2.8, 19)\,\mathrm{s}^{-1}$.

\begin{figure}[tb]
    \centering
    \includegraphics[width=0.55\columnwidth]{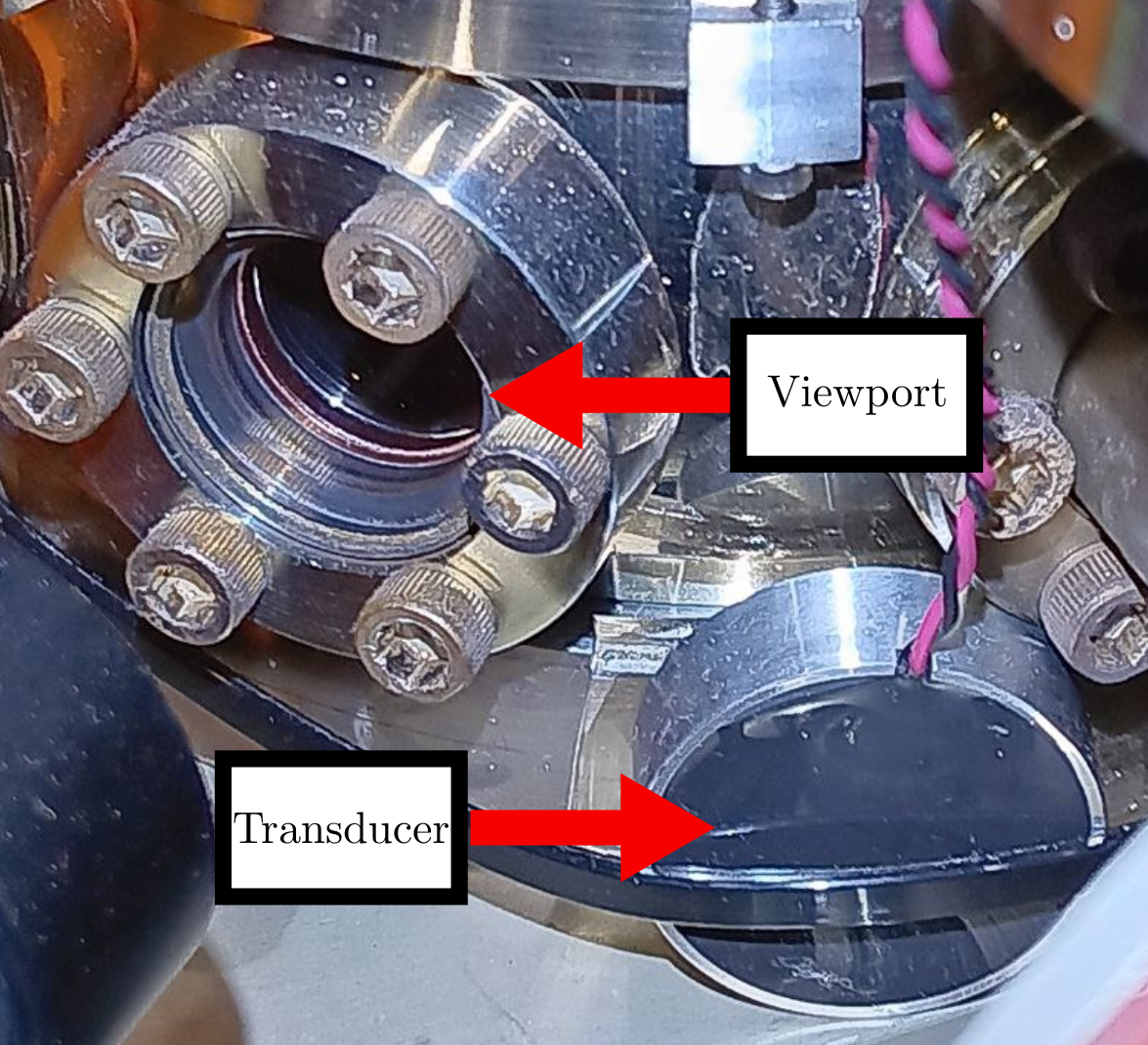}
    \includegraphics[width=0.4\columnwidth]{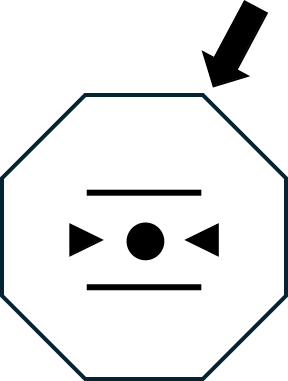}
    \caption{Experimental apparatus. Left: Transducer mounted on the outer wall of the vacuum chamber. Right: Orientation of the transducer oscillation axis relative to the trap denoted by the arrow. The filled circle, triangles, and lines correspond to the ion, end caps, and trap rods respectively.}
    \label{fig:transducer}
\end{figure}

Each experiment begins by preparing the ion near the motional ground state using electromagnetically induced-transparency cooling~\cite{eit1,eit2} followed by resolved-sideband cooling, yielding a mean phonon occupation of $\bar n\approx0.1$.  The ion is then prepared in the electronic ground-state $|g\rangle$ via optical pumping, with typical state preparation and detection errors of $1.2\%$.

Mechanical oscillations are generated by a piezoelectric transducer (Part number: H2KLPY11000600, Unictron Technologies Corporation) attached to the outer walls of the vacuum chamber (Fig.~\ref{fig:transducer}, left). A direct digital synthesizer controls the drive frequency and amplitude. The transducer is positioned such that the oscillation axis has components along all three directions of ion motion (Fig.~\ref{fig:transducer}, right).

\begin{figure}[!t]
    \centering
    \includegraphics[width=\columnwidth]{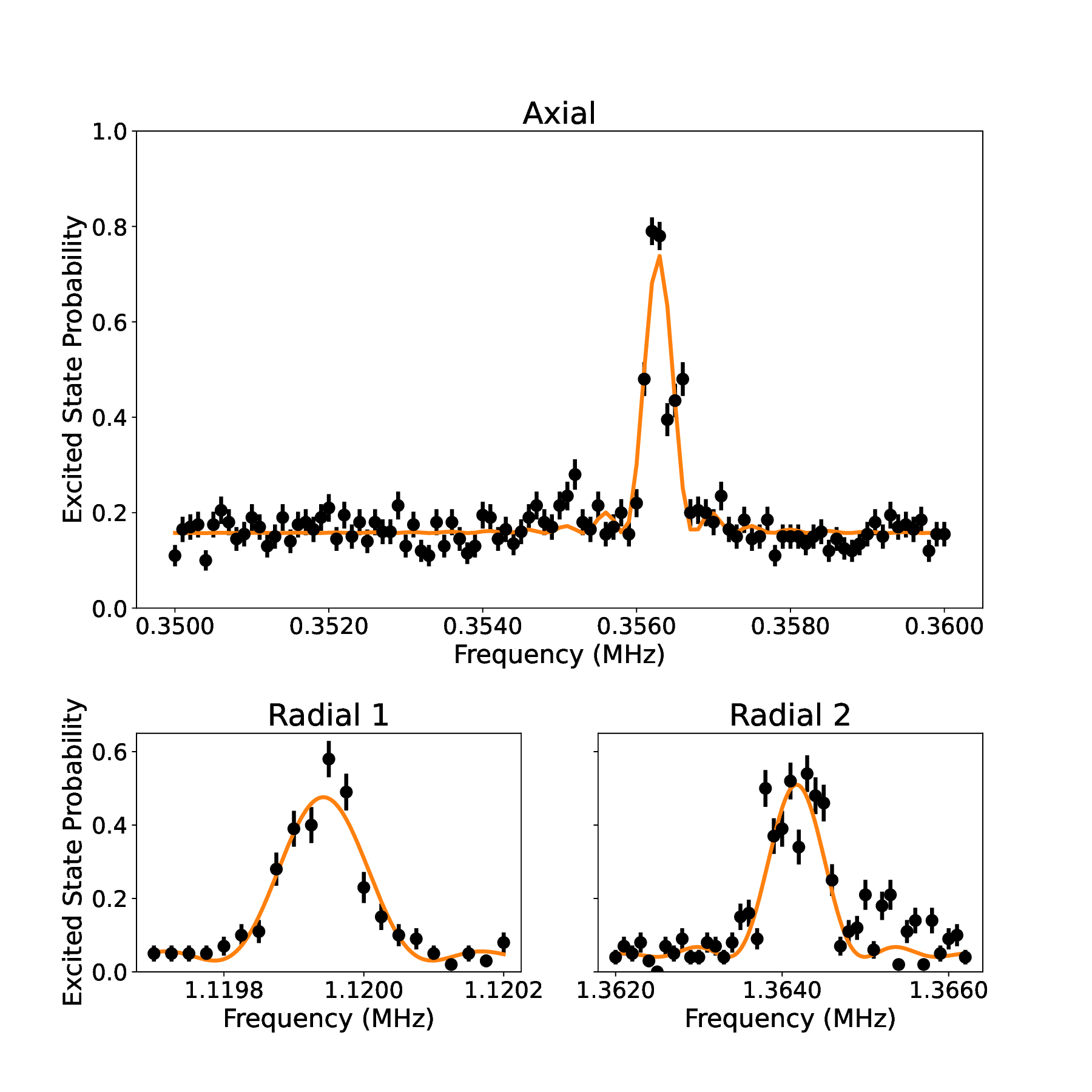}
    \caption{ 
    Top: Excited-state population versus transducer drive frequency for axial mode. Error bars indicate statistical uncertainty from 100 repeated measurements, and the orange curve is a sinc-squared fit. A clear peak is observed, corresponding to the axial mode frequency. The nonzero background is attributed to imperfect cooling of the mode.
    Bottom: Same measurements for the two radial modes.}
    \label{fig:fig2}
\end{figure}

\begin{figure*}[tbh]
    \centering
    \includegraphics[width=0.99\textwidth]{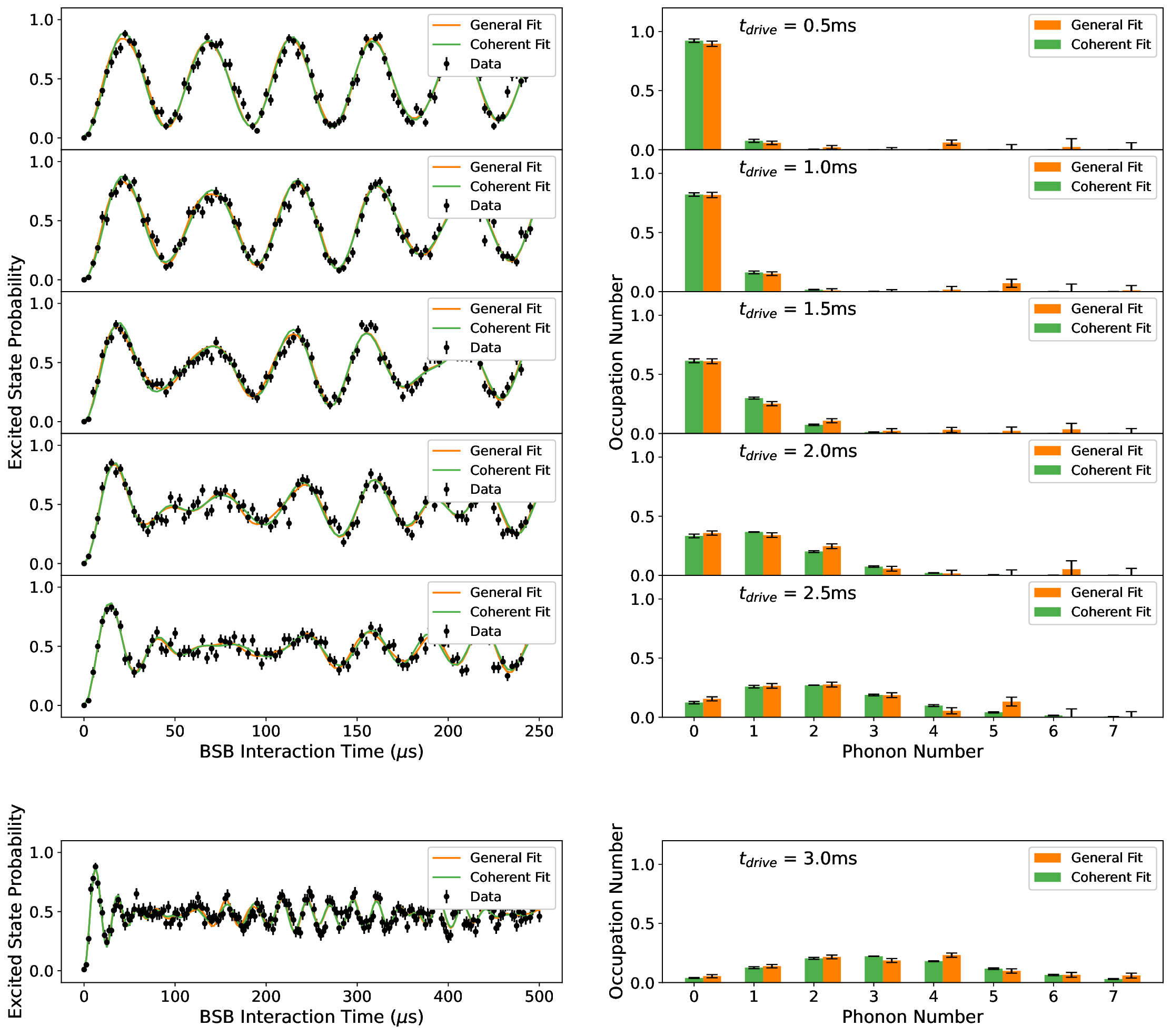}
    \caption{Extracted phonon populations for varying transducer drive durations, at a fixed drive amplitude of $122\, {\rm mV_{\rm pp}}$. Left: Excited-state population as a function of blue-sideband interaction time. Right: Reconstructed populations of the first eight Fock states. Orange lines and markers show a model-free reconstruction in which each Fock state population is treated as independent fit parameters. Green lines and markers show a coherent state model with the populations following a Poisson distribution. The close agreement indicates that the dynamics generated by the mechanical drives are well-described as coherent displacements.}
    \label{fig:fig4}
\end{figure*}

\section{Results}
\subsection{Demonstration}
We first verify the proposed coupling scheme using the following sequence. After state initialization, the transducer is driven at a fixed amplitude of $615$ $\mathrm{mV}_{\rm pp}$ for $5$ ms while the drive frequency is scanned. After a 0.5 ms ringdown delay, we probe the motional state with a red-sideband pulse of duration $t_\pi/2$, followed by internal-state readout. A nonzero population in $\ket{e}$ indicates that the transducer drive has coupled to a motional mode and transferred population out of the motional ground state.

The measured response is plotted in Fig.~\ref{fig:fig2}. A pronounced feature is observed near the axial mode frequency at 0.356284(8) MHz, verifying that mechanical vibrations do couple to the ion motion. Signals are also present for the radial mode frequencies 1.119942(3) MHz and 1.364173(15) MHz, for the drive amplitude of 1.23 $\mathrm{V}_{\rm pp}$ and drive times of 10~ms and 4~ms respectively. However, significant frequency fluctuations of the radial modes hindered further experimentation. We, therefore, focus on the axial mode in the subsequent analysis.

\subsection{Dynamics}
Mechanical vibrations induced by the transducer cause the trap structure, and hence the trapping potential, to oscillate. If dephasing and heating of the motional state are negligible on the experimental timescale, this process is expected to implement a displacement operation that transforms an initial vacuum into a coherent state~\cite{glauber1963}. To test this expectation, we resonantly drive the axial mode with the transducer after state preparation, varying the drive amplitude and duration. We then probe the resulting motional state with a blue-sideband pulse followed by internal-state readout. Figure~\ref{fig:fig4} shows representative data for a drive amplitude of $122\,\mathrm{mV}_{\rm pp}$.

The phonon-number distribution is extracted by fitting the time-dependent excited-state population to
\begin{equation}
    \label{eqn:bsb_population}
    P_e(t) = \frac{A}{2} \bigg( B - \sum_{n=0}^N p_n \cos(\Omega_{n, n+1} t) e^{-\lambda (n+2)^{0.7} t} \bigg)
\end{equation}
where $A$ and $B$ are amplitude and offset parameters, $p_n$ is the population of the $n^{\text{th}}$ Fock state, $\Omega_{n,n+1}$ is the blue-sideband Rabi frequency, and $\lambda$ accounts phenomenologically for decoherence~\cite{Meekhof1996}. We apply two models to the data in Fig.~\ref{fig:fig4}. The first (in orange) is model-free and makes no assumptions about the underlying phonon-number distribution. The second (in green) constrains the populations to a Poisson distribution, as expected for a coherent state. Good agreement between the two reconstructions is consistent with the generation of a coherent state.

\begin{figure}[t]
    \centering
    \includegraphics[width=\columnwidth]{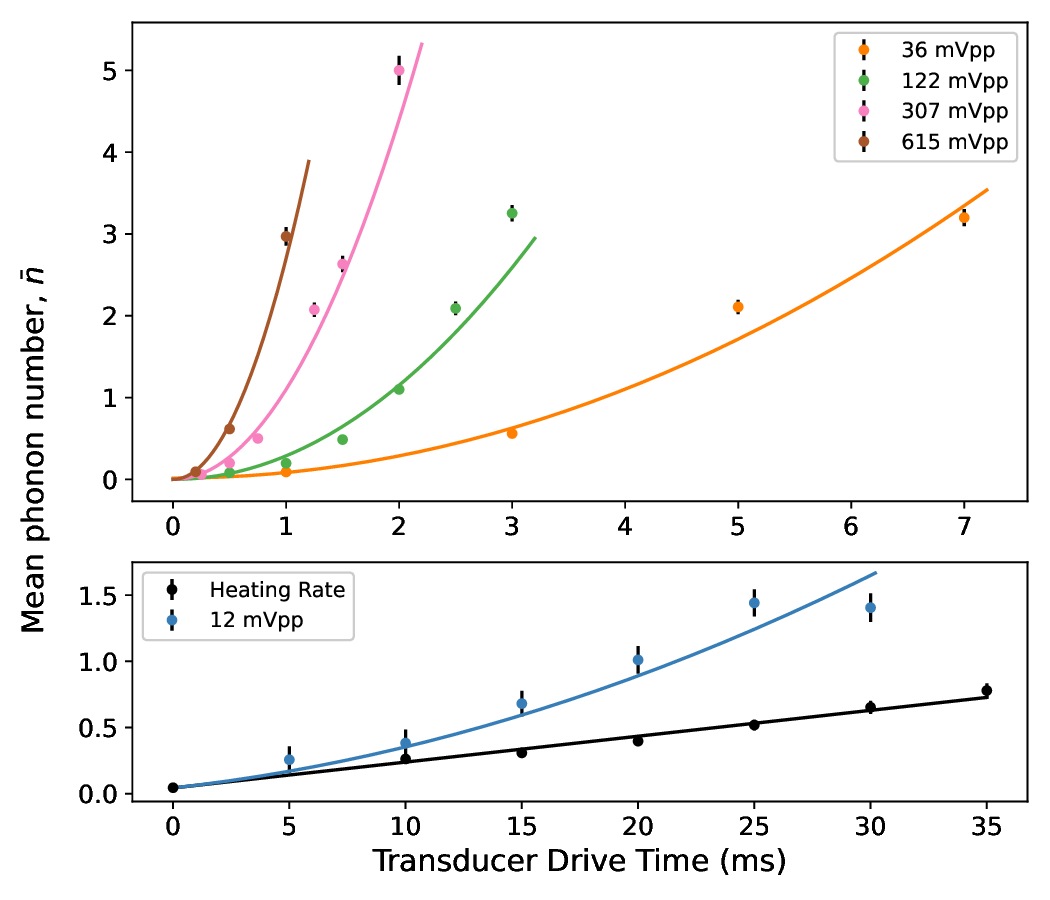}
    \caption{Top: Mean phonon number ($\bar{n}$) extracted from coherent state fits versus transducer drive time for several drive amplitudes. Some error bars are smaller than the markers. Quadratic fits capture the increase of $\bar{n}$ with drive time, consistent with resonant displacement dynamics.
    Bottom: Heating measurements and experimental data for the 12~$\mathrm{mV}_{\rm pp}$ transducer drive amplitude. For the longest 12 mV\(_{\rm pp}\) drives, heating appreciably changes the phonon-number distribution. We therefore use model-free population reconstruction of the first five Fock states to estimate phonon numbers for the 12~$\mathrm{mV}_{\rm pp}$ data. We assume thermal states for the heating measurements. 
    The fit uses the model given by Eq.\ref{eq:quantum_model_solved}. }
    \label{fig:fig5}
\end{figure}

To examine the system's time evolution, Fig.~\ref{fig:fig5} shows the mean phonon number as a function of transducer drive duration for several drive amplitudes. The observed quadratic growth of $\bar n$ with time, in contrast to the linear growth caused by heating, is characteristic of resonant coherent displacement.

\subsection{Estimation of Trap Displacement}
We use a Lindblad master equation to estimate the displacement amplitude of the trapping potential in the presence of dephasing and heating.  Mechanical vibration of the trap shifts the center of the trapping potential and applies a force $f(t)=-M\omega^2 d \cos(\omega t)$ on the ion. Here $\omega$ is the motional angular frequency, $M$ is the ion mass, and $d$ is the trap-center displacement amplitude. This interaction is described by the Hamiltonian $ H = -\hat x f(t)$. When the frequency of trap vibrations is on resonance with the mode frequency, this Hamiltonian can be written in the rotating frame of the harmonic oscillator as 
\begin{equation}
    H = \hbar g (a + a^{\dagger}),
\end{equation}
where $a$ and $a^\dagger$ are the annihilation and creation operators respectively, $g = \omega d / 4 z_0$, and $z_0=\sqrt{\frac{\hbar}{2M\omega}}$ is the spatial extent of the ground state wave function. 

The equations of motion for the expectation values of displacement, $\alpha = \langle a \rangle$, and average phonon number, $\bar n = \langle a^\dagger a\rangle$, including Markovian motional heating and frequency dephasing described by the Lindblad operators $L_1 = \sqrt{2\gamma} a^\dagger a$, $L_2 = \sqrt{\Gamma} a$, and  $L_3 = \sqrt{\Gamma} a^\dagger$~ \cite{Turchette_PRA_2000,MotionalErrors_PhysRevA}, take the form
\begin{eqnarray}
    \frac{d\alpha}{dt} &=& -\gamma \alpha - i g \nonumber \\ 
    \frac{d \bar n}{dt} &=& i g (\alpha - \alpha^*) + \Gamma,
\end{eqnarray}
where $\gamma$ is the pure dephasing rate and $\Gamma$ is the motional heating rate. For $\alpha(0)=0$ and $\bar n(0)=n_0$, the solutions are 
\begin{eqnarray} \label{eq:quantum_model_solved}
    \alpha(t) & = & -i \frac{g}{\gamma}(1 - e^{-\gamma t}) \nonumber \\
    \bar n(t) & = &  n_0 + \Gamma t + \frac{2 g^2}{\gamma^2}(\gamma t - 1 + e^{-\gamma t}). 
\end{eqnarray}
In the limit of negligible dephasing ($\gamma t  \ll 1$) the amplitude grows linearly, and the drive-induced contribution to $\bar n$ increases quadratically with time
\begin{equation}
  \alpha = -i g t\, ; \quad\quad\quad  \bar n(t) = n_0 + \Gamma t + g^2 t^2 
\end{equation}
consistent with the ideal classical oscillator model~\cite{Landau1976Mechanics}. In the opposing limit, when the drive time is much longer than the oscillator dephasing time ($\gamma t  \gg 1$), the mean phonon number grows linearly with time 
\begin{equation} \label{eq:quantum_coherent_drive}
     \bar n(t) \simeq n_0 +  \left( \Gamma + \frac{2 g^2}{\gamma} \right) t 
\end{equation}
and the external drive effectively increases the heating of the motional mode.  

The lower panel of Fig. \ref{fig:fig5} shows the data for the heating and mean phonon number measurements for the transducer drive amplitude of 12 $\mathrm{mV}_{\rm pp}$. The linear fit of the heating data determines the heating rate $\Gamma = 19(1) {\rm \; s^{-1}}$ and initial phonon occupation $n_0 = 0.04(2)$. Independently measured decay of the motional $(|0\rangle + |1\rangle)/\sqrt{2}$ state coherence using a Ramsey-type experiment has characteristic $1/e$ time of $\tau_c = 37(4) {\rm \; ms}$ almost entirely consistent with the measured heating rate~\cite{Turchette_PRA_2000}. We estimate additional dephasing rate $\gamma < 2 {\rm \; s^{-1}}$. Fitting Eq. (\ref{eq:quantum_model_solved}) to the 12 $\mathrm{mV}_{\rm pp}$ data with the parameters $\Gamma$ and $n_0$ determined from the heating rate measurements yields a trap displacement amplitude
\begin{equation}
    d =  \frac{4 g z_0}{\omega } \approx 6(1)\times10^{-13}\,\mathrm{m} = 0.6(1)\,\mathrm{pm}.
\end{equation}
showing that the ion is sensitive to sub-picometer amplitudes of trap vibration. The quoted uncertainty comes mainly from the variations of reconstructed phonon-number distribution with the Fock number cutoff. 

\section{Conclusion}
We have demonstrated direct coupling between externally driven mechanical vibrations and the motional dynamics of a single trapped ion. By driving a transducer mounted on the outer wall of the vacuum chamber, we observed a clear resonance response of the axial mode. The extracted phonon populations and their time evolution are both consistent with a displacement operation acting on the motional state. These measurements show that the ion is sensitive to sub-picometer motion of the trap.

This configuration represents an intentionally strong perturbation: most trapped-ion systems operate on vibration-isolated platforms, and high-frequency mechanical motion is further attenuated as it propagates through the surrounding structures and air. However, low frequency modes, such as the axial center-of-mass mode of a long ion chain~\cite{Cetina_Axial_heating_2022}  may be more susceptible to external mechanical vibrations. Our results show that mechanically induced motion can have measurable effects on trapped-ion motional states. With suitable extensions, this sensitivity could also be relevant to proposed searches for high-frequency gravitational waves~\cite{mhz_gravitational_waves_2026,gravitational_waves_ions_2026}.

Finally, mechanical driving offers a practical route to motional mode frequency characterization. Because the method does not rely on optical driving of sideband interactions, it can avoid light shift related offsets that arise in standard optical measurements and may reduce the need for longer Ramsey-style calibration sequences. In contrast to direct application of resonant electric signals to the trap, this technique does not require any compromise in the design of low pass filters for trap electrodes~\cite{NIST_Inderect_cooling_2024}. Such measurements of the unperturbed mode frequency could be used to calibrate pulse sequences for entangling gates and to implement mode-frequency stabilization protocols.

\bibliographystyle{unsrt}
\bibliography{references}

\begin{acknowledgments}
 This project is supported by the National Research Foundation, Singapore through the National Quantum Office, hosted in A*STAR, under its Centre for Quantum Technologies Funding Initiative (S24Q2d0009) and Quantum Engineering Programme (W21Qpd0208). 
\end{acknowledgments}

\section*{Author Declarations}
\subsection*{Conflict of Interest}
The authors report no conflicts of interest.
\subsection*{Author Contributions}
\noindent\textbf{Mu Young Kim:} Conceptualization (equal); Formal analysis (equal); Investigation (lead); Methodology (lead); Visualization (lead); Writing – original draft (lead); Writing – review \& editing (equal). 
\noindent\textbf{H. C. J. Gan:} Conceptualization (equal); Formal analysis (supporting); Investigation (lead); Methodology (lead);  Visualization (supporting); Writing – review \& editing (equal). 
\noindent\textbf{Nigel Benjamin Lee Junsheng:} Formal analysis (supporting); Visualization (supporting); Writing – review \& editing (equal). 
\textbf{Dzmitry Matsukevich:} Conceptualization (equal); Formal analysis (supporting), Funding acquisition (lead); Project administration (lead); Supervision (lead); Writing – review \& editing (equal).
\\
\section*{Data Availability Statement}
The data that support the findings of this study are available from the corresponding author upon reasonable request.

\end{document}